\documentclass[nofootinbib,aps,10pt,twocolumn,superscriptaddress]{revtex4-1}

\usepackage[dvips]{graphicx}

\usepackage{array,amsmath,amsthm,feynmf,mathtools}
\usepackage{mathtools}
\usepackage{epsfig}
\usepackage{graphicx}
\usepackage{color}
\usepackage{slashed}
\usepackage{amssymb}

\DeclareMathSymbol{\ScriptD}{\mathord}{symbols}{228}

\DeclareMathSymbol{\DoubleStruckOne}{\mathord}{operators}{177}

\DeclareMathSymbol{\DoubleStruckG}{\mathord}{operators}{231}

\DeclareMathOperator{\Tr}{Tr}

\newcommand{\packageX}{\emph{Package}-{\tt\bfseries X}}

\newcommand{\comm}[1]{{\fontsize{10}{10}\selectfont$\mathtt{#1}$}}

\makeatletter
\let\saved@underbrace\underbrace
\renewcommand*\underbrace[1]{\@ifnextchar_{\ub@with{#1}}{\ub@without{#1}}}
\def\ub@with#1_#2{\mathpalette\underbrace@i{{#1}{_{#2}}}}
\newcommand*\ub@without[1]{\mathpalette\underbrace@i{{#1}{}}}
\newcommand*\underbrace@i[2]{\underbrace@ii#1#2}
\newcommand*\underbrace@ii[3]{\saved@underbrace{#1#2}#3}
\makeatother

\begin{document}

\title{\emph{Package}-X: A \emph{Mathematica} package for the analytic calculation of one-loop integrals}
\author{Hiren H. Patel}
\email{hiren.patel@mpi-hd.mpg.de}
\affiliation{Particle and Astro-Particle Physics Division \\
Max-Planck Institut fuer Kernphysik {\rm{(MPIK)}} \\
Saupfercheckweg 1, 69117 Heidelberg, Germany}
\begin{abstract}
%
\packageX, a \emph{Mathematica} package for the analytic computation of one-loop integrals dimensionally regulated near 4 spacetime dimensions is described.  {\packageX} computes arbitrarily high rank tensor integrals with up to three propagators, and gives compact expressions of UV divergent, IR divergent, and finite parts for any kinematic configuration involving real-valued external invariants and internal masses.  Output expressions can be readily evaluated numerically and manipulated symbolically with built-in \emph{Mathematica} functions.  Emphasis is on evaluation speed, on readability of results, and especially on user-friendliness.  Also included is a routine to compute traces of products of Dirac matrices, and a collection of projectors to facilitate the computation of fermion form factors at one-loop.  The package is intended to be used both as a research tool and as an educational tool.\\[10mm]
\noindent\textbf{Program summary}\\[2mm]
\emph{Program title:} Package-X\\[2mm]
\emph{Program obtainable from:} CPC Program Library, Queen's University, Belfast, N. Ireland, \emph{or} {\tt http://packagex.hepforge.org}\\[2mm]
\emph{Licensing provisions:} Standard CPC license, {\tt http://cpc.cs.qub.ac.uk/licence/licence.html}\\[2mm]
\emph{Programming language:} Mathematica (Wolfram Languange)\\[2mm]
\emph{Operating systems:} Windows, Mac OS X, Linux (or any system supporting Mathematica 8.0 or higher)\\[2mm]
\emph{RAM required for execution:} 10 MB, depending on size of computation\\[2mm]
\emph{Vectorised/parallelized?:} No\\[2mm]
\emph{Nature of problem:} Analytic calculation of one-loop integrals in relativistic quantum field theory for arbitrarily high-rank tensor integrals and any kinematic configuration of real-valued external invariants and internal masses.\\[2mm]
\emph{Solution method:} Passarino-Veltman reduction formula, Denner-Dittmaier reduction formulae, and two new reduction algorithms described in the manuscript.\\[2mm]
\emph{Restrictions:} One-loop integrals are limited to those involving no more than three propagator factors.\\[2mm]
\emph{Unusual features:} Includes rudimentary routines for tensor algebraic operations and for performing traces over Dirac gamma matrices.\\[2mm]
\emph{Running Time:} 5ms to 10s for integrals typically occurring in practical computations; longer for higher rank tensor integrals.



\end{abstract}
\maketitle
\section{Introduction}
Many packages are available to assist with the evaluation of one-loop integrals that appear in higher order calculations of perturbative quantum field theory.  The most widely used ones are the \emph{Mathematica} packages \textsc{FeynCalc}\cite{Mertig:1990an}, \textsc{FormCalc}\cite{Hahn:1998yk} and the Fortran program \textsc{Golem95}\cite{Binoth:2008uq}.  These packages compute one-loop integrals using the Passarino-Veltman reduction algorithm\cite{Passarino:1978jh} (\textsc{FeynCalc} and \textsc{FormCalc} also feature a collection of routines designed to streamline the numerical computation of a differential cross section; as such, they do substantially more than to simply compute one-loop integrals).  

Nevertheless \textsc{FeynCalc} falls short in that it gives results of one-loop computations in terms of basis scalar functions which cannot be evaluated on their own.  Instead, it is up to the user to supply their analytical forms from an external source, or to link them to yet another package (such as \textsc{FF}\cite{vanOldenborgh:1990yc}, \textsc{LoopTools}\cite{Hahn:1998yk}, or \textsc{OneLOop}\cite{vanHameren:2010cp}).

Moreover, one-loop integrals have many more applications than to calculate cross sections and decay rates.  Examples are the computation of ultraviolet counterterms, pole positions, residues, Peskin-Takeuchi oblique parameters, electromagnetic moments, \emph{etc}.  Many of these applications require the calculation of Feynman integrals at singular kinematic points such as at physical thresholds or zero external momenta.  Since the Passarino-Veltman reduction algorithm typically breaks down at these points, it is nearly impossible to obtain results with \textsc{FeynCalc} or \textsc{FormCalc} (\textsc{Golem95} can give numerical results).  But, it is also at these points where compact analytic expressions exist. 

Although smaller-scale packages are available that are designed around a particular application (such as \textsc{lool}\cite{Ilakovac:2014yka} and \textsc{ant}\cite{Angel:2013hla}), there is no general-purpose software that gives analytic results to one-loop integrals for all kinematic configurations.  In this regard, {\packageX} serves to fill this gap.

\begin{widetext}
{\packageX} calculates dimensionally regulated ($d=4-2\epsilon$) rank-$P$ one-loop tensor integrals of the form
\begin{equation}
T_N^{\mu_1\ldots\mu_P}(p_1,\ldots,p_N;m_0,m_1\ldots,m_N)=
\mu^{2\epsilon}\!\!\int\!\! \frac{d^d k}{(2\pi)^d} \frac{k^{\mu_1} \cdots k^{\mu_P}}{[k^2\!-\!m_0^2\!+\!i\varepsilon][(k\!+\!p_1)^2\!-\!m_1^2\!+\!i\varepsilon]\cdots[(k\!+\!p_N)^2\!-\!m_N^2\!+\!i\varepsilon]}\,,
\end{equation}
\end{widetext}
with up to $N=3$ denominator factors, and finds compact analytic expressions for arbitrary configurations of external momenta $p_i$ and real-valued internal masses $m_i$.  The functional paradigm of the \emph{Wolfram Language} together with the supplementary trace-taking routines included in {\packageX} allows one to compute an entire one-loop diagram at once.  All output is ready for numerical evaluation and symbolic manipulation with \emph{Mathematica}'s internal functions.


  This article details the technical aspects of \packageX, and assumes familiarity in the use of the package.  The application files are found at the Hepforge webpage \mbox{\tt http://packagex.hepforge.org}, where the software will be maintained and periodically updated.  Included among the package files is a tutorial that provides an introduction, and a complete set of documentation files that becomes embedded with the \emph{Wolfram Documentation Center} upon installation which provides details and examples of all functions defined in \packageX.

\section{Structure and Design of package}
\begin{figure}\vspace{0mm}
\includegraphics[width=5.3cm]{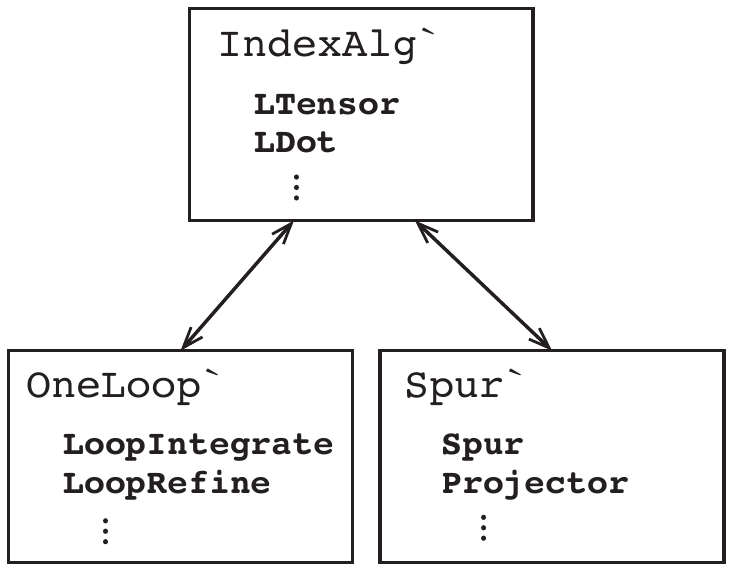}
\caption{Organization of functions into \emph{contexts} as defined in \packageX.}\label{fig:organization}
\end{figure}

The subroutines in this package belong to one of three \emph{Mathematica contexts} organized as in Fig. \ref{fig:organization}.  The module \comm{IndexAlg`} contains the rudimentary tensor-algebraic routines and serves as the backbone of \packageX.  \comm{OneLoop`} contains the algorithms and look-up tables for the computation of one-loop integrals, and \comm{Spur`} includes the algorithms to perform traces over products of Dirac matrices and contains a catalog of fermion form factor projectors.

The basic {\packageX} workflow for the computation of a one-loop integral consists of three steps:
\begin{enumerate}
\item Call \comm{LoopIntegrate} to carry out the covariant tensor decomposition (section \ref{sec:LoopIntegrate}).
\item Apply on-shell conditions and other kinematic relations with \emph{Mathematica}'s built-in functions \comm{ReplaceAll} ({\tt /.}) and \comm{Rules} (\comm{\rightarrow}).
\item Call \comm{LoopRefine} to convert coefficient functions into explicit expressions (section \ref{sec:LoopRefine}).
\end{enumerate}
The reasoning behind the three-step design is as follows: kinematic configurations of external invariants $p_i.p_j$ and internal masses $m_i$ relevant to many physical applications occur at singular points of one-loop integrals, such that if they were applied \emph{after} obtaining the general expressions, errors like \comm{0/0} or \comm{0\times ln(0)} would inevitably occur.  To avoid such errors and to facilitate the generation of compact results, \comm{LoopRefine} uses algorithms depending critically on the kinematic configuration supplied by the user \emph{beforehand}.

Two other supplementary functions are provided to streamline computations involving fermions: 
\begin{itemize}
\item \comm{Spur} (\emph{German for `trace'}) computes traces of Dirac matrices that may appear in the numerators of one-loop integrals (section \ref{sec:Spur}).
\item \comm{Projector} is used to project fermion self-energy and vertex form factors out of the loop integrals (section \ref{sec:Projector}).
\end{itemize}
The algorithms used by these functions are detailed in the aforementioned sections below.

\begin{widetext}
\section{{\tt LoopIntegrate}: Covariant tensor decomposition}\label{sec:LoopIntegrate}
The evaluation of an integral is initiated with \comm{LoopIntegrate}, which carries out its covariant tensor decomposition in terms of scalar coefficient functions.  For example (omitting the $+i\varepsilon$),
\begin{align}
\nonumber\text{\comm{LoopIntegrate[k_\mu k_\nu k_\rho, k, p1,m0,m1]:}}&\\
\big({\textstyle\frac{i}{16\pi^2}}\big)^{\!-1}\mu^{2\epsilon} \!\!\int\!\! \frac{d^d k}{(2\pi)^d} \frac{k^\mu k^\nu k^\rho}{[k^2-m_0^2][(k+p_1)^2-m_1^2]}
&\longrightarrow\{[p_1][g]\}^{\mu\nu\rho} B_{001} + \{[p_1]^3\}^{\mu\nu\rho} B_{111}\,,\\[5mm]
\nonumber\text{\comm{LoopIntegrate[k_\mu k_\nu, k, p1,p2,m0,m1,m2]:}}&\\
\nonumber\big({\textstyle\frac{i}{16\pi^2}}\big)^{\!-1}\mu^{2\epsilon} \!\!\int\!\!  \frac{d^d k}{(2\pi)^d} \frac{k^\mu k^\nu}{[k^2-m_0^2][(k+p_1)^2-m_1^2][(k+p_2)^2-m_2^2]}
&\longrightarrow\\
&\hspace{-2cm}\{[g]\}^{\mu\nu} C_{00} + \{[p_1]^2\}^{\mu\nu} C_{11} + \{[p_1][p_2]\}^{\mu\nu} C_{12} 
+ \{[p_2]^2\}^{\mu\nu} C_{22}
\end{align}
\end{widetext}
Here, $B_{001}$, $B_{111}$, $C_{00}$ \emph{etc.} are coefficient functions that depend only on Lorentz invariants, $p_i.p_j$ and $m_i$.  Note that as indicated in the left hand sides, an overall constant $(\frac{i}{16\pi^2})$ is factored out of the natural integration measure $\mu^{2\epsilon}\frac{d^dk}{(2\pi)^d}$ to simplify the output.  Each coefficient function multiplies a totally symmetric tensor, denoted $\{\ldots\}^{\mu\ldots}$ in the notation of \cite{Denner:2005nn}, containing products of external momentum four-vectors $p_i^\mu$ and the metric tensor $g^{\mu\nu}$.   These tensors are generated by a {\packageX} internal function (inside \comm{IndexAlg`}),
which utilizes \emph{Mathematica}'s built-in function \comm{Permutations}.  The time to generate the corresponding symmetric tensors grows factorially with the rank of tensor integrals.  

For integrals with high powers of contracted loop momenta, such as
\begin{equation}
\mu^{2\epsilon} \int \frac{d^d k}{(2\pi)^d} \frac{k^\alpha k^\beta\,(k.k)^5}{[k^2-m_0^2][(k+p_1)^2-m_1^2]}\,,
\end{equation}
it is necessary to obtain explicit expressions of self-contracted symmetrized high-rank tensors like $\{[p_1]^6[g]^3\}^{\alpha\beta\mu \mu \nu\nu\rho\rho \sigma\sigma\lambda\lambda}$.  It would be wasteful to first generate the totally symmetric high-rank tensors, only to subsequently contract indices down to lower-rank symmetric tensors.  Instead, the contraction formulae
\begin{multline}
(p_k)_{\mu_1}\{[p_1]^{n_1}\cdots[p_N]^{n_N}[g]^r\}^{\mu_1\ldots\mu_P}\\
=\sum_{\ell=1}^N p_k\cdot p_\ell \{[\hat{p}_\ell] [p_1]^{n_1}\cdots[p_N]^{n_N}[g]^r \}^{\mu_2\ldots \mu_P}\\
+(n_k+1)\{[p_k][p_1]^{n_1}\cdots[p_N]^{n_N}[g]^{r-1}\}^{\mu_2\ldots \mu_P}
\end{multline}
\begin{multline}
g_{\mu_1 \mu_2}\{[p_1]^{n_1}\cdots[p_N]^{n_N}[g]^r\}^{\mu_1\ldots\mu_P}\\
=\sum_{i,j}^N p_i \cdot p_j \{[\hat{p}_i] [\hat{p}_j] [p_1]^{n_1}\cdots[p_N]^{n_N}[g]^r \}^{\mu_3\ldots \mu_P}\\
+\bar{\delta}_{r,0}(d+P-2+\sum_k^N n_k)\{[p_1]^{n_1}\cdots[p_N]^{n_N}[g]^{r-1}\}^{\mu_3\ldots \mu_P}\,,
\end{multline}
are employed to carry out the self-contractions symbolically before converting any remaining symmetric tensors with free indices into explicit forms in terms of $p_i^\mu$ and $g^{\mu\nu}$.  The time to construct self-contracted tensors in this way is reduced to follow a power law.

\section{{\tt LoopRefine}: Reduction to elementary functions}\label{sec:LoopRefine}
Once the covariant decomposition is made, and any on-shell or kinematic conditions are applied, the final step is to feed the results into \comm{LoopRefine}, which replaces the coefficient functions with explicit expressions.  The basic algorithm followed by \comm{LoopRefine} is as follows:
\begin{description}
\item[\sc{Step} 1] For each coefficient function (\comm{pvA}, \comm{pvB}, \comm{pvb}, or \comm{pvC}) encountered by \comm{LoopRefine}, symbols 
for internal masses are recorded (for {\sc Step 4}), and the appropriate reduction routine (see corresponding subsections below) is called.

\item[\sc{Step} 2] The reduction of $C$ functions for more general kinematic configurations end with the scalar function $C_0$.  If the $C_0$ function is IR-divergent or has an explicit form that is sufficiently compact (as controlled by the option \comm{ExplicitC0}), the explicit form is substituted.

\item[\sc{Step} 3] All instances of the spacetime dimension $d$ is replaced by $4-2\epsilon$, and \emph{Mathematica}'s built-in function \comm{Series} is called to keep the leading terms in the $\epsilon$ expansion.  UV-divergences appear as $1/\epsilon$ poles, and IR-divergences appear as $1/\epsilon$ and/or $1/\epsilon^2$ poles.

\item[\sc{Step} 4] Combine and simplify logarithms, organize the expression by the logarithms, and group the 't Hooft parameter $\mu^2$-dependent logarithm with the $1/\epsilon$ pole in the expression (see section \ref{sec:ieps}).

\end{description}


In the following subsections, the algorithms and accompanying formulae used by \comm{LoopRefine} to reduce the coefficient functions are summarized.  It should be noted that nearly all algorithms are drawn from the 2005 paper by Denner and Dittmaier \cite{Denner:2005nn}, and will be referenced henceforth as [DD].  The only formulae not taken directly from their paper are those for the auxiliary $b^\xi$ functions in Section \ref{sec:reductionAuxB} (which is only a slight modification of the reduction formulae for $B$ functions), and those of two additional algorithms for the reduction of $C$ functions in special kinematic configurations (\emph{Cases 2} and \emph{4} in section \ref{sec:reductionC}).

\subsection{Reduction of $A$ and $B$ functions}\label{sec:ABreduction}
The Passarino-Veltman coefficient $A$ functions are simple enough to be obtained by direct integration (eqn 3.4 of [DD]): 
\begin{equation}
A_{\underbrace{0 \ldots 0}_{2r}}(m_0) = \frac{(m_0^2)^{r+1}}{2^r(r+1)!}\Big(\frac{1}{\bar\epsilon}+\ln(\frac{\mu^2}{m_0^2})+H_{r+1}\Big)\,,
\end{equation}
where $1/\bar{\epsilon} = 1/\epsilon-\gamma_E+\ln(4\pi)$, and $H_n$ is the $n^\text{th}$ harmonic number.

The $B_{0\ldots0\,1\ldots1}$ functions, with at least one pair of $00$ indices are obtained iteratively in terms of those with fewer number of $00$ indices using (eqn 4.5 of [DD]):
\begin{multline}
B_{\underbrace{0\ldots 0}_{2r} \underbrace{1\ldots 1}_n}(p^2;m_0,m_1) = \frac{-1}{2(n+1)}\big[(-1)^{n+1} A_{\underbrace{0\ldots0}_{\mathclap{2(r-1)}}}(m_1)\\
+(p^2-m_1^2+m_0^2)B_{\underbrace{0\ldots 0}_{\mathclap{2(r-1)\enspace}} \underbrace{1\ldots 1}_{\mathclap{\enspace n+1}}}(p^2;m_0,m_1)\\
+2p^2 B_{\underbrace{0\ldots 0}_{\mathclap{2(r-1)\enspace}} \underbrace{1\ldots 1}_{\mathclap{\enspace n+2}}}(p^2;m_0,m_1)\big]\,,\qquad r\geq1
\end{multline}
Then the $B_{1\ldots1}$ integrals (with no $00$ index pairs) are obtained by explicit integration over the single Feynman parameter in (\ref{eq:FeynParB}).  Results are given in (eqn 4.8 of [DD]), but the form that tends to generate most compact expressions is
\begin{widetext}
\begin{align}
B_{\underbrace{1\ldots 1}_n}(p^2;m_0,m_1) &=
\nonumber \frac{(-1)^n}{n+1}\Big[\frac{1}{\bar\epsilon} + \ln\big(\frac{\mu^2}{m_1^2}\big) + \sum_{k=0}^n \frac{2}{n+1} \sum_{j=0}^{\mathclap{\left \lfloor{\frac{n-k}{2}}\right \rfloor}} \binom{n-k}{j}\left(\frac{p^2+m_0^2-m_1^2}{2p^2}\right)^{n-k-2j}\left(\frac{\lambda(p^2,m_0^2,m_1^2)}{4(p^2)^2}\right)^j\\
\nonumber &\qquad - \sum_{k=0}^{\mathclap{\left \lfloor{\frac{n-1}{2}}\right \rfloor}} \binom{n+1}{2k}\left(\frac{p^2+m_0^2-m_1^2}{2p^2}\right)^{n+1-2k}\left(\frac{\lambda(p^2,m_0^2,m_1^2)}{4(p^2)^2}\right)^k \ln\left(\frac{m_0^2}{m_1^2}\right)\\
&\qquad +\sum_{k=0}^{\left \lfloor{\frac{n}{2}}\right \rfloor} \binom{n+1}{2k+1} \left(\frac{p^2+m_0^2-m_1^2}{2p^2}\right)^{n-2k}\left(\frac{\lambda(p^2,m_0^2,m_1^2)}{4(p^2)^2}\right)^k \Lambda(p^2;m_0,m_1)
\Big]\,.
\end{align}
Here $\lambda(a,b,c)=a^2+b^2+c^2$ is the K\"{a}ll\'{e}n function, implemented as \comm{Kallen\lambda[a,b,c]}, and $\Lambda(p^2;m_0,m_1)$ is the abbreviation
\begin{equation}
\Lambda(p^2;m_0,m_1) = \frac{\sqrt{\lambda(p^2,m_0^2,m_1^2)}}{p^2}
\ln\Big(\frac{2 m_0 m_1}{-p^2+m_0^2+m_1^2-\sqrt{\lambda(p^2,m_0^2,m_1)}}+i\varepsilon\Big)\,,
\end{equation}
implemented as \comm{DiscB[s,m0,m1]}.  In order to access $B_{1\ldots1}(p^2;m_0,m_1)$ at its singular points, a limiting procedure would need to be made at runtime in order to avoid errors such as \comm{0/0} or \comm{0\times ln(0)}.  While \emph{Mathematica}'s function \comm{Limit} can eventually generate an expression, computation time is long, and output expressions are always unwieldy.  Instead, a catalog of explicit expressions (also obtained by direct integration) of $B_{1\ldots1}$ at all its singular points (see Table \ref{tab:ListOfCases}) is included in the source code.
They may be accessed directly within {\packageX} using \comm{LoopRefine[pvB[0}$,n,s,m_0,m_1$\comm{]]}.
\end{widetext}
\subsection{Reduction of auxiliary $b^\xi$ functions}\label{sec:reductionAuxB}
In covariant gauges, the propagator for massless vector fields 
\begin{equation}
i \tilde{D}^{\mu\nu}(k)=\frac{-i}{k^2}\Big[g^{\mu\nu}-(1-\xi)\frac{k^\mu k^\nu}{k^2}\Big]\,,
\end{equation}
contains a gauge term that leads to an additional factor in the denominator of one-loop integrals. 
{\packageX} can handle such propagators inside bubble integrals, with the coefficient functions given by the auxiliary Passarino-Veltman $b^\xi$ functions \cite{Bardin:1999ak}.
For example,
\begin{multline}\nonumber
\big({\textstyle\frac{i}{16\pi^2}}\big)^{\!-1}\mu^{2\epsilon}\int \frac{d^d k}{(2\pi)^d}\frac{k^\mu k^\nu k^\rho}{[k^2]^2 [(k+p)^2-m^2]}\\
=\{[p][g]\}^{\mu\nu\rho} b_{001}^\xi + \{[p]^3\}^{\mu\nu\rho} b_{111}^\xi\,.
\end{multline}
The reduction formulae for these functions essentially mirror those for the standard $B$ functions.  Auxiliary $b^\xi_{0\ldots0\,1\ldots1}$ functions with at least one pair of $00$ indices are iteratively determined in terms of functions with fewer $00$ index pairs using
\begin{multline}
b^\xi_{\underbrace{0\ldots 0}_{2r} \underbrace{1\ldots 1}_n}(p^2;m) = \frac{-1}{2(n+1)}\big[B_{\underbrace{0\ldots0}_{\mathclap{2(r-1)}}}(p^2;0,m)\\
+(p^2-m^2)b^\xi_{\underbrace{0\ldots 0}_{\mathclap{2(r-1)\enspace}} \underbrace{1\ldots 1}_{\mathclap{\enspace n+1}}}(p^2;m)\\
+2p^2 b^\xi_{\underbrace{0\ldots 0}_{\mathclap{2(r-1)\enspace}} \underbrace{1\ldots 1}_{\mathclap{\enspace n+2}}}(p^2;m)\big]\,,\qquad r\geq1\,,
\end{multline}
and the $b^\xi_{1\ldots1}$ functions with no $00$ index pairs are obtained by direct integration over the single Feynman parameter in (\ref{eq:FeynParBaux}).  The integral is finite if $n\geq1$, with the result
\begin{multline}\nonumber
b^\xi_{\underbrace{1\ldots 1}_n}(p^2;m) = \\
 \frac{(-1)^{n-1}}{p^2}\Big[-\frac{1}{n}+\sum_{k=1}^{n-1}\frac{1}{n-k}\frac{m^2}{p^2-m^2}\left(\frac{p^2-m^2}{p^2}\right)^k\\
+ \frac{m^2}{p^2-m^2}\left(\frac{p^2-m^2}{p^2}\right)^n\ln\left(\frac{m^2}{m^2-p^2}+i\varepsilon\right)\Big]\,.
\end{multline}
If $n=0$ (a case that is not met in practice since the gauge part of the spin-1 propagator guarantees two powers of momenta in the numerator), the auxiliary $b^\xi$ function is IR-divergent.

Explicit expressions at the various singular points of $b_{1\ldots1}^\xi$ (see Table \ref{tab:ListOfCases}) are included in the {\packageX} source code.

\subsection{Reduction of $C$ functions}\label{sec:reductionC}
The reduction of coefficient $C$ functions is significantly complicated by its numerous singular points.  Although the standard Passarino-Veltman reduction algorithm is applicable at almost all points (\emph{Case 1} below), different formulae are needed to handle the various singular cases (\emph{Cases 2 -- 6}). \comm{LoopRefine} identifies the nature of the kinematic configuration and applies the appropriate reduction method.   

\emph{Cases 1}, \emph{3}, \emph{5} and \emph{6} are taken from [DD].  Note that since the emphasis of [DD] is on numerical stability and not on generating analytic expressions, the algorithms presented there do not automatically give compact expressions.  The algorithm under \emph{Case 2} is new, and while technically it is covered by \emph{Case 1}, it leads to more compact expressions.  Furthermore, an algorithm to handle the reduction at physical thresholds (applied in \emph{Case 3} below) is not completely covered by [DD].  This gap is filled by the formulae under \emph{Case 4}.

 The arguments of the coefficient $C$ functions are ordered differently in {\packageX} as compared to those used by other authors.  See Appendix \ref{app:conventions} for details.

In the reduction formulae below, the following kinematic abbreviations are used (which differ slightly from [DD] by numeric factors):
\begin{equation}
\begin{array}{ll}
f_j = p_j^2 - m_j^2 + m_0^2\,, & \hspace{-5mm}j = \{1,\,2\}\\[4mm]
Z = \begin{pmatrix} p_1^2 & p_1.p_2 \\ p_2.p_1 & p_2^2\end{pmatrix} & \text{(Gramian matrix)}\\[4mm]
q^2 = p_1^2 + p_2^2 -2 p_1.p_2\\[4mm]
\det Z = \frac{1}{4}\lambda(q^2,\,p_1^2,\,p_2^2)\\[4mm]
\tilde{Z} = \begin{pmatrix}p_2^2 & -p_1.p_2 \\ -p_1.p_2 & p_1^2 \end{pmatrix} & \text{(cofactor matrix)}\\[4mm]
\tilde{X}_{0j} = \begin{pmatrix} p_2^2 f_1 - p_1.p_2 f_2 \\ -p_1.p_2 f_1 + p_1^2 f_2 \end{pmatrix} & j = \{1,\,2\}
\end{array}
\end{equation}
Furthermore, hatted indices on coefficient functions (\emph{e.g.} $B_{\hat{k}0\ldots0\, 1\ldots1}$) indicate the removal of those indices.  Coefficient $B$ functions derived by canceling denominators from three-point integrals are abbreviated by
\begin{align}
B_{...}(\hat{D}_1) =& B_{...}(p_2^2; m_0, m_2)\\
B_{...}(\hat{D}_2) =& B_{...}(p_1^2; m_0, m_1)\,.
\end{align}
If the denominator $(k^2 - m_0^2)^{-1}$ independent of an external momentum vector is cancelled, a shifted form of the $B$ function is used:
\begin{multline}
B_{\underbrace{0\ldots0}_{2r}\underbrace{1\ldots1}_{n_1}\underbrace{2\ldots2}_{n_2}}(\hat{D}_0) =\\
 (-1)^{n_1} \sum_{j=0}^{n_1}\binom{n_1}{j}B_{\underbrace{0\ldots0}_{2r}\underbrace{1\ldots1}_{\mathclap{n_2+j}}}(q^2; m_1, m_2)\,.
\end{multline}
Whenever $n_1 > n_2$ the invariance property
\begin{equation}
B_{\underbrace{0\ldots0}_{2r}\underbrace{1\ldots1}_{n_1}\underbrace{2\ldots2}_{n_2}}(\hat{D}_0) = B_{\underbrace{0\ldots0}_{2r}\underbrace{1\ldots1}_{n_2}\underbrace{2\ldots2}_{n_1}}(\hat{D}_0)\Big|_{m_1\leftrightarrow m_2}
\end{equation}\\
is used to keep the number of terms in the sum to a minimum.  \emph{Cases 2} and \emph{4} require expressions for the $B$ functions with the number of $00$ index pairs continued to $r = -1$.  Details of this function are found in Appendix \ref{App:algB}.

Finally, formulae for \emph{Cases 1, 3} and \emph{5} below contain explicit dependence on spacetime dimension $d=4-2\epsilon$ appearing in denominators of certain prefactors.  In the course of reduction, the $\mathcal{O}(\epsilon)$ part multiplying any lower coefficient functions combines with their UV $1/\epsilon$ poles\footnote{
For the argument that they are only of UV origin (and not IR), see the argument in Sec. 5.8 of [DD]\\}, and gives rise to finite polynomials in kinematic variables.  Although this can be automatically handled by \comm{Series} at {\sc Step 3}, the reduction algorithm performs much faster if these polynomials are explicitly supplied.  They are obtained by integration over the Feynman parameters as described at the end of Appendix \ref{sec:FeynPar}.

\begin{widetext}

\subsubsection*{Case 1: $\det Z \neq 0$}

At non-singular kinematic configurations with $\det Z\neq 0$, the original \cite{Passarino:1978jh} Passarino-Veltman reduction formula is used (eqns 5.10, 5.11 of [DD]):
\begin{equation}
\left\{
\begin{aligned}
C_{\underbrace{0\ldots 0}_{2r} \underbrace{1\ldots 1}_{n_1} \underbrace{2\ldots 2}_{n_2}} &= \frac{1}{2 \det Z}\sum_{k=1}^2 \tilde{Z}_{jk} \Big[\delta_{n_k,\delta_{jk}}B_{\underbrace{0\ldots 0}_{2r} \underbrace{1\ldots 1}_{\mathclap{n_{\bar{k}-\delta_{\bar{k}1}}}}}(\hat{D}_k)-B_{\underbrace{0\ldots 0}_{\mathclap{2r}} \underbrace{1\ldots 1}_{\mathclap{n_1-1}} \underbrace{2\ldots 2}_{\mathclap{n_2}}}(\hat{D}_0)\\[-2mm]
&\hspace{4cm}-f_k C_{\underbrace{0\ldots 0}_{\mathclap{2r}} \underbrace{1\ldots 1}_{\mathclap{n_1-1}} \underbrace{2\ldots 2}_{\mathclap{n_2}}}(\hat{D}_0)-2(n_k-\delta_{jk})C_{\hat{k} \underbrace{0\ldots 0}_{\mathclap{2r+2}} \underbrace{1\ldots 1}_{\mathclap{n_1-1}} \underbrace{2\ldots 2}_{\mathclap{n_2}}}\Big]\,,& n_1\geq 1\\
C_{\underbrace{0\ldots 0}_{2r}} &= \frac{1}{2(d-4+2r)}\Big[B_{\underbrace{0\ldots 0}_{\mathclap{2r-2}}}(\hat{D}_0) + 2m_0^2 C_{\underbrace{0\ldots 0}_{\mathclap{2r-2}}1}+f_2 C_{\underbrace{0\ldots 0}_{\mathclap{2r-1}}2}\Big]\,,& r\geq1
\end{aligned}
\right.
\end{equation}
where $\bar k = \begin{cases} 1\,,\enspace & k=2\\ 2\,, & k=1\end{cases}$. In the first equation, $j=1$ is taken, although the choice $j=2$ would give equivalent results.  If $n_1=0$ with $n_2>0$, then the relation (\ref{eq:invarianceOfC}) is used and the first equation is applied.
\subsubsection*{Case 2: Ellis-Zanderighi triangle 6}
Coefficient $C$ functions for which arguments are $(m_0^2, s, m_2^2; m_2, 0, m_0)$---or an equivalent permutation thereof---are already covered by \emph{Case 1}.  However, final expressions obtained from it tend not to give the most compact formulae for this kinematic configuration.  More compact formulae are obtained by directly integrating over the Feynman parameters in (\ref{eq:FeynParC}); see Appendix \ref{app:SpecialCaseDerivation} for derivation.  It is of note that the corresponding scalar function $C_0$ is the IR-divergent three-point function, `triangle 6', as classified by Ellis and Zanderighi \cite{Ellis:2007qk}.  In Eqs.~(\ref{eq:ez6A}) and (\ref{eq:ez6B}), it is assumed that at least one of $r$, $n_1$ or $n_2$ is nonzero.
\begin{equation}\label{eq:ez6A}
\left\{
\begin{gathered}
\shoveleft{C_{\underbrace{0\ldots 0}_{2r} \underbrace{1\ldots 1}_{n_1} \underbrace{2\ldots 2}_{n_2}}(m_0^2, s, m_2^2; m_2, 0, m_0)} = \\[-2mm]
\shoveright{\hspace{3.8cm}\frac{(-1)^{n_1}}{2}\frac{n_1 ! (n_2 +2r - 1)!}{(n_1+n_2+2r)!}\big(1+2\epsilon(H_{n_1+n_2+2r}-H_{n_2+2r-1})\big)B_{\underbrace{0\ldots0}_{2r-2}\underbrace{1\ldots1}_{n_2}}(s;m_0,m_2)\,,}\\[-1mm]
\shoveright{\text{$n_2\neq0$ or $r\neq0$}}\\
\shoveleft{C_{\underbrace{1\ldots 1}_{n_1}}(m_0^2, s, m_2^2; m_2, 0, m_0)} = \\[-2mm]
\shoveright{(-1)^{n_1}\Big[C_0(m_0^2, s, m_2^2; m_2, 0, m_0)-\frac{1}{2}\Big(H_{n_1}+\epsilon(H_{n_1}^2 + H_{n_1}^{(2)})\Big)B_{\underbrace{0\ldots 0}_{-2}}(s;m_0,m_2)\Big]}
\end{gathered}
\right.
\end{equation}
where $H_n^{(r)}$ is the $n^\text{th}$ harmonic number of order $r$.  If the arguments take the form $(s, m_0^2, m_2^2; 0, m_2^2, m_0)$, then the identity (\ref{eq:invarianceOfC}) is applied, and the equations above are valid.

A different formula is needed if the off-shell momentum $s$ is in the third position:
\begin{multline}\label{eq:ez6B}
C_{\underbrace{0\ldots 0}_{2r} \underbrace{1\ldots 1}_{n_1} \underbrace{2\ldots 2}_{n_2}}(m_2^2, m_0^2, s; m_0, m_2, 0) =
 \frac{(-1)^{n_2}}{2}\frac{1}{n_1+n_2+2r}\sum_{k=0}^{n_2}\binom{n_2}{k}\big(1+\frac{2\epsilon}{n_1+n_2+2r}\big) B_{\underbrace{0\ldots 0}_{\mathclap{2r-2}} \underbrace{1\ldots 1}_{\mathclap{n_1+k}}}(s;m_0,m_2)
\end{multline}
To apply Eqs. (\ref{eq:ez6A}) and (\ref{eq:ez6B}) above, explicit forms of the scalar $B$ function with the number of $00$ index pairs taken to $r=-1$ is occasionally needed.  These functions are discussed in Appendix \ref{App:algB}.

\subsubsection*{Case 3: $\det Z = 0$ but $\tilde {X}_{0j}\neq 0$}
With $\det Z=0$, the primary reduction formulae are rearranged to give: (eqns 5.38 and 5.40 of [DD])
\begin{equation}
\left\{
\begin{aligned}
C_{\underbrace{0\ldots 0}_{2r}} &= \frac{1}{d+2r-3}\Big(B_{\underbrace{0\ldots0}_{2r-2}}(\hat{D}_0) - m_0^2 C_{\underbrace{0\ldots 0}_{2r-2}}\Big) + 
\frac{1}{2(d+2r-3)\tilde{Z}_{kl}}\sum_{n,m=1}^{2} \Big(\delta_{km}\delta_{nl}-\delta_{kl}\delta_{nm}\Big)\\
&\hspace{-1.5cm}\times\Big\{\sum_{j=1}^2 Z_{nj}\Big[(1-\delta_{mj}) B_{\underbrace{0\ldots 0}_{2r-2} 1}(\hat{D}_m)-B_{j \underbrace{0\ldots 0}_{2r-2}}(\hat{D}_0)\Big]+\frac{1}{2}f_m \Big[-B_{\underbrace{0\ldots 0}_{2r-2}}(\hat{D}_n)+B_{\underbrace{0\ldots 0}_{2r-2}}(\hat{D}_0) + f_n C_{\underbrace{0\ldots 0 }_{2r-2}}\Big]\Big\} \qquad\mathclap{r>0}\\[2mm]
C_{\underbrace{0\ldots 0}_{2r} \underbrace{1\ldots 1}_{n_1} \underbrace{2\ldots 2}_{n_2}} &= \frac{1}{\tilde{X}_{0j}}\sum_{k=1}^2 \tilde{Z}_{jk}\Big(\delta_{n_k 0} B_{\underbrace{0\ldots 0}_{2r} \underbrace{1\ldots 1}_{\mathclap{n_{\bar{k}}}}}(\hat{D}_k) - B_{\underbrace{0\ldots 0}_{2r} \underbrace{1\ldots 1}_{n_1} \underbrace{2\ldots 2}_{n_2}}(\hat{D}_0) - 2n_k C_{\hat{k}00\underbrace{0\ldots 0}_{2r} \underbrace{1\ldots 1}_{n_1} \underbrace{2\ldots 2}_{n_2}}\Big)
\end{aligned}
\right.
\end{equation}
The value of $j$ chosen (1 or 2) is the one for which the corresponding $\tilde X_{0j}$ is non-vanishing.  If both elements are vanishing, then \emph{Case 4} is applied.  Note that the second relation is valid even when either $n_1=0$ or $n_2=0$.  In particular, when $r=n_1=n_2=0$ the final term in that relation vanishes, and leads to the reduction of the scalar $C_0$ function in terms of scalar $B_0$ functions.
\subsubsection*{Case 4: vanishing $\det Z$ and $\tilde {X}_{0j}$}
When the physical threshold (corresponding to $\tilde X_{0j}$ = 0 for both $j=\{1,2\}$) coincides with the boundary of the physical region ($\det Z = 0$), then \emph{Cases 1}---\emph{3} are inapplicable. For this kinematic configuration, the reduction formulae in [DD] eqns (5.49) and (5.53) can be used provided at least one element of 
\begin{equation*}
\tilde X_{ij} =\begin{pmatrix}
4m_0^2 p_2^2 - f_2^2 & -2m_0^2(p_1^2+p_2^2-q^2)+f_1f_2 \\ -2m_0^2(p_1^2+p_2^2-q) + f_1f_2 & 4m_0^2 p_1^2 - f_1^2
\end{pmatrix}
\end{equation*}
 is non-vanishing.  However, no reduction methods are presented in [DD] that are valid when all four elements of $\tilde X_{ij}$ are vanishing, because an expansion around that point is not known\footnote{A. Denner, \emph{private correspondence}}.  This exceptional configuration is needed for the computation of elastic form factors at zero momentum such as electron $g-2$.  To fill this gap, a new set of reduction formulae are used that is valid regardless of the form of $\tilde X_{ij}$, provided at least one of $p_1^2$, $p_2^2$ or $q^2$ is non-vanishing.  These formulae are derived in Appendix \ref{sec:DerivationOfCase4}.

If $p_2^2\neq0$,
\begin{multline}\label{eq:reductionAtThres1}
C_{\underbrace{0\ldots 0}_{2r} \underbrace{1\ldots 1}_{n_1} \underbrace{2\ldots 2}_{n_2}} = \frac{(-1)^{n_1+n_2}}{2}\sum_{j=0}^{n_2}\binom{n_2}{j}\alpha^{n_2-j}\bigg\{\frac{n_1!(n_2-j)!}{(n_1+n_2-j+1)!}\sum_{k=0}^j\Bigg[ \binom{j}{k}(-\alpha)^{j-k}(-1)^k B_{\underbrace{0\ldots 0}_{2r-2} \underbrace{1\ldots 1}_{k}}(\hat{D}_1)\Bigg] \\
+\sum_{k=0}^{n_1}\frac{(-1)^{n_2}}{n_2-j+k+1}\binom{n_1}{k}\Big[(1-\alpha)^{j+1}(-1)^{n_2}B_{\underbrace{0\ldots 0}_{\mathclap{2r-2\enspace}} \underbrace{1\ldots 1}_{\mathclap{\enspace n_2+k+1}}}(\hat{D}_0)
-(-\alpha)^{j+1}B_{\underbrace{0\ldots 0}_{\mathclap{2r-2\enspace}} \underbrace{1\ldots 1}_{\mathclap{\enspace n_2+k+1}}}(\hat{D}_2)
\Big]
\bigg\}\,,
\end{multline}
where $\alpha = -q^2 + p_1^2 + p_2^2/(2p_2^2)$.

If $p_2^2 = 0$, then $\det Z=0$ implies $q^2 = p_1^2$, and the formula
\begin{equation}\label{eq:reductionAtThres2}
C_{\underbrace{0\ldots 0}_{2r} \underbrace{1\ldots 1}_{n_1} \underbrace{2\ldots 2}_{n_2}} = \frac{(-1)^{n_1+1}}{2(n_2+1)}\sum_{k=0}^{n_1} \binom{n_1}{k} B_{\underbrace{0\ldots 0}_{\mathclap{2r-2\enspace}} \underbrace{1\ldots 1}_{\mathclap{\enspace n_2+k+1}}}(\hat{D}_2)
\end{equation}
is used.  If $p_1^2=p_2^2=q^2=0$, then these formulae are inapplicable and \emph{Case 5} is needed.  Note that when $r=0$ in either (\ref{eq:reductionAtThres1}) or (\ref{eq:reductionAtThres2}), the $B$ functions continued to $r=-1$ are needed (see Appendix \ref{App:algB}).
\subsubsection*{Case 5: All elements of $Z$ vanishing}
If all external invariants are vanishing $p_1^2=p_2^2=q^2=0$, then the following are applied (eqns 5.62 and 5.63 of [DD]):
\begin{equation}
\left\{
\begin{aligned}
C_{\underbrace{0\ldots 0}_{2r} \underbrace{1\ldots 1}_{n_1} \underbrace{2\ldots 2}_{n_2}} &= \frac{1}{d+2(n_1+n_2 + r-1)}\Big[B_{\underbrace{0\ldots 0}_{2r-2} \underbrace{1\ldots 1}_{n_1} \underbrace{2\ldots 2}_{n_2}}(\hat{D}_0) + m_0^2 C_{\underbrace{0\ldots 0}_{2r-2} \underbrace{1\ldots 1}_{n_1} \underbrace{2\ldots 2}_{n_2}}\Big]\,,& r\geq1\\
C_{\underbrace{1\ldots 1}_{n_1} \underbrace{2\ldots 2}_{n_2}} &= \frac{1}{f_k}\Big[\delta_{n_k 0} B_{\underbrace{1\ldots 1}_{n_{\bar k}}}(\hat{D}_k)-B_{\underbrace{1\ldots 1}_{n_1} \underbrace{2\ldots 2}_{n_2}}(\hat{D}_0)-2n_k C_{\hat{k}00 \underbrace{1\ldots 1}_{n_1} \underbrace{2\ldots 2}_{n_2}}\Big]\end{aligned}
\right.
\end{equation}
In the second equation, the index $k$ is chosen such that $f_k$ is non-vanishing.  As in \emph{Case 3}, the second relation is valid for vanishing $n_1$ or $n_2$, and is used to reduce the scalar $C_0$ function to $B_0$ functions for $n_1=n_2=0$.
\subsubsection*{Case 6: All elements of $Z$ and $f_k$ vanishing}
Finally, if also the $f_k$'s are vanishing, the following formulae are used (eqns 5.71 and 5.72 of [DD]):
\begin{equation}
\left\{
\begin{aligned}
C_{\underbrace{0\ldots 0}_{2r} \underbrace{1\ldots 1}_{n_1} \underbrace{2\ldots 2}_{n_2}} &= \frac{-1}{2(n_k+1)}B_{k\underbrace{0\ldots 0}_{2r-2} \underbrace{1\ldots 1}_{n_1} \underbrace{2\ldots 2}_{n_2}}(\hat{D}_0)\,,& r\geq 1\\
C_{\underbrace{1\ldots 1}_{n_1} \underbrace{2\ldots 2}_{n_2}} &= \frac{1}{m_0^2}\Big[(d+2n_1+2n_2)C_{00 \underbrace{1\ldots 1}_{n_1} \underbrace{2\ldots 2}_{n_2}} - B_{\underbrace{1\ldots 1}_{n_1} \underbrace{2\ldots 2}_{n_2}}(\hat{D}_0)\Big]
\end{aligned}
\right.
\end{equation}
Equivalent results are obtained for $k = 1$ or $2$ in the first equation.  The choice $k=1$ is used in \packageX.  Note that when $r=n_1=n_2=0$, these relations together permit the reduction of the scalar $C_0$ function in terms of scalar $B_0$ functions.
\end{widetext}

\begin{table*}
\rule{\textwidth}{1pt}
\begin{flushleft}
\begin{tabular}{p{3.4cm}p{3.4cm}p{3.4cm}p{3.7cm}}
\multicolumn{2}{l}{\emph{$B$-functions} --- Section \ref{sec:ABreduction}}\\[1mm]
\hline
$B_{1\ldots1}(0;0,0)$ & $B_{1\ldots1}(p^2;0,0)$ & $B_{1\ldots1}(m_0^2; m_0, 0)$ & $B_{1\ldots1}(p^2; m_0, 0)$\\
$B_{1\ldots1}(m_0^2;m_0,m_0)$ & $B_{1\ldots1}(0; m_0, m_0)$ & $B_{1\ldots1}(0; m_0,m_1)$ & $B_{1\ldots1}((m_0+m_1)^2;m_0,m_1)$ \\
$B_{1\ldots1}(0;0,m_1)$ & $B_{1\ldots1}(m_1^2;0,m_1)$ & $B_{1\ldots1}(p^2; 0, m_1)$ & $B_{1\ldots1}((m_0-m_1)^2;m_0,m_1)$ \\
$B_{1\ldots1}(0;m_0,0)$
\end{tabular}\\[0.5cm]
\begin{tabular}{p{3.4cm}p{3.4cm}p{3.4cm}p{3.7cm}}
\multicolumn{2}{l}{\emph{$B$-functions with $r=-1$} --- Appendix \ref{App:algB}}\\[1mm]
\hline
$B_{1\ldots1}(0;0,0)$ & $B_{1\ldots1}(p^2;0,0)$ & $B_{1\ldots1}(m_0^2; m_0, 0)$ & $B_{1\ldots1}(p^2; m_0, 0)$\\
$B_{1\ldots1}(0;0,m_1)$ & $B_{1\ldots1}(0; m_0, m_0)$ & $B_{1\ldots1}(0; m_0,m_1)$ & $B_{1\ldots1}((m_0+m_1)^2;m_0,m_1)$ \\
$B_{1\ldots1}(0;m_0,0)$ & $B_{1\ldots1}(m_1^2;0,m_1)$ & $B_{1\ldots1}(p^2; 0, m_1)$ & $B_{1\ldots1}((m_0-m_1)^2;m_0,m_1)$
\end{tabular}\\[0.5cm]
\begin{tabular}{p{3.5cm}p{3.5cm}p{4cm}}
\multicolumn{2}{l}{\emph{Auxiliary $b^\xi$-functions} --- Section \ref{sec:reductionAuxB}}\\[1mm]
\hline
$b^\xi_{1\ldots1}(0,0)$ & $b^\xi_{1\ldots1}(p^2,0)$\\
$b^\xi_{1\ldots1}(0,m)$ & $b^\xi_{1\ldots1}(m^2,m)$
\end{tabular}\\[0.5cm]

\begin{tabular}{p{3.5cm}p{4cm}p{4.5cm}p{5cm}}
\multicolumn{4}{l}{\emph{Scalar C-functions} --- Section \ref{sec:C0function} }\\[1mm]
\hline
$C_0(0,0,0;0,0,0)$ & 			$C_0(0,0,q^2;0,m_0,m_0)$	&	$C_0(m_0^2,0,q^2;0,0,m_0)$		&	$C_0(0,p_2^2,q^2;m_2,0,0)$\\
$C_0(0,0,q^2;0,0,0)$ &			$C_0(0,0,q^2;0,m_1,m_0)$	&	$C_0(0,m_2^2,q^2;m_2,0,0)$	&	$C_0(p_1^2,0,q^2;m_2,m_1,m_0)$\\
$C_0(0,0,m_2^2;m_2,0,0)$ &		$C_0(0,0,q^2;m_0,m_0,m_0)$	&	$C_0(m_0^2,0,m_2^2;m_2,0,m_0)$		&	$C_0(m_0^2,m_0^2,q^2;0,0,m_0)$\\
$C_0(0,0,q^2;m_2,0,0)$ &		$C_0(0,0,q^2;m_2,m_0,m_0)$	&	$C_0(p_1^2,p_2^2,q^2;m_2,m_1,m_0)$		&	$C_0(m_0^2,p_2^2,m_0^2;m_0,0,m_0)$\\
$C_0(0,0,q^2;0,0,m_0)$ &		$C_0(0,0,q^2;m_2,m_1,m_0)$	&	$C_0(0,m_0^2,q^2;0,m_0,m_0)$			&	$C_0(m_0^2,p_2^2,m_2^2;m_2,0,m_0)$\\
$C_0(0,0,m_0^2;m_0,m_0,m_0)$ &	$C_0(p_1^2,0,q^2;0,0,0)$ 	&	$C_0(0,p_2^2,q^2;m_0,m_0,m_0)$		&	$C_0(p_1^2,p_2^2,q^2;0,0,0)$
\end{tabular}
\end{flushleft}
\caption{Special kinematic cases of the Passarino-Veltman coefficient functions $B$, $b^\xi$ and $C$ for which explicit expressions are included in the source file {\tt OneLoop.m}.  Further information for these functions is found in the indicated sections}\label{tab:ListOfCases}
\rule{\textwidth}{1pt}
\end{table*}

\section{The scalar $C_0$ function: Analytic expressions and numerical implementation}\label{sec:C0function}
The algorithms for the reduction of coefficient $C$-functions for which $\det Z \neq 0$ (\emph{Cases 1} and \emph{2} in the previous section) end with the UV-finite scalar function $C_0(p_1^2,p_2^2,q^2,m_2,m_1,m_0)$.  To complete the computation of the one loop integral and to make the final output useable, the scalar function must be replaced.  For this purpose, a complete catalog of analytic expressions for $C_0$ where $\det Z \neq 0$---each one obtained by direct integration---is included in the source file (see Table \ref{tab:ListOfCases}).  
Many such expressions are scattered throughout the literature.  The general formula with non-zero kinematic variables appears in \cite{Hooft:1978xw}.  All IR-divergent three-point formulae are given in \cite{Ellis:2007qk}, 
and some special cases appear in unpublished notes \cite{JorgeRomao:2004a}.

Although a complete catalog of analytic expressions of $C_0$ is available, not all cases are automatically substituted by \comm{LoopRefine} at {\sc Step 2}.  The functions that are substituted are only those that are IR-divergent (to faithfully display the $1/\epsilon$ poles in the final output), and those for which a sufficiently simple/compact expression is known.  For more complicated finite cases, \comm{LoopRefine} simply outputs\footnote{If the explicit analytic form \emph{is} desired, the option \mbox{\comm{ExplicitC0 \rightarrow All}} can be supplied to \comm{LoopRefine}.} \mbox{\comm{pvC0[}$\ldots$\comm{]}}, with the function itself implemented numerically, (summarized below).  The reason for this design choice is as follows:

Firstly, in cases for which no simple form is known, the general formula \cite{Hooft:1978xw} in terms of 12 dilogarithms would have to be given.  This expression for $C_0$ alone would occupy a very large part of the output overwhelming the remainder of the expression, thus defeating the original purpose of producing \emph{compact} expressions.  Secondly, the dilogarithm function is computationally very expensive.  When numerics are required, a brute-force evaluation of all the dilogarithms is grossly inefficient, leading to excessively slow numerical evaluations.

The main features of the code for the rapid numerical evaluation of the three-point scalar function for real masses and external momenta are as follows:
\begin{itemize}
\item The imaginary part of $C_0$ in the physical region (defined by $\lambda(q^2, p_1^2, p_2^2)>0$) is obtained by applying Cutkosky's rule, and with a straightforward continuation into the unphysical region (defined by $\lambda(q^2, p_1^2, p_2^2)<0$) \cite{Fronsdal:1963zz, Lucha:2006vc}.  Its computation requires the evaluation of a single logarithm. 

\item  The real part of $C_0$ requires evaluations of only the real part (in the physical region) or only the imaginary part (in the unphysical region) of the dilogarithm, but not both.  Calling \comm{PolyLog} would lead to needless computation of both parts by the \emph{Mathematica} Kernel.  Following \cite{Osacar:1995aa}, the real and imaginary parts of the dilogarithm function are implemented separately.

\item  For the real part of $C_0$ in the physical region, the $+i\varepsilon$ prescription is irrelevant (since it influences only the imaginary part which is anyway evaluated using Cutkosky's rule).  Then, either the arguments of the 12 dilogarithms come in complex-conjugate pairs (for which the real part of the dilogarithms are identical and are added reducing the number of dilog evaluations), or the arguments are purely real (for which the real parts of the dilogarithms are rapidly evaluated using real arithmetic).

\item The code is compiled to the \emph{Wolfram Virtual Machine} (using \comm{Compile}), leading to a substantial boost in computation speed.
\end{itemize}

In the physical region, up to a 200-fold increase in speed is achieved compared to brute-force \emph{Mathematica} evaluation by the Kernel.  In the unphysical region, up to a 20-fold increase in speed is obtained.  If the option \comm{CompilationTarget\!\rightarrow\!``C"} to \comm{Compile} is set, its performance rivals that of the Fortran implementation in {\sc LoopTools}, with {\packageX} generating results approximately twice as fast.

\section{Handling the $+i\varepsilon$ prescription and simplifying logarithms}\label{sec:ieps}
The $+i\varepsilon$ prescription appearing in the denominators of propagator functions enforce causality in the time-ordered Green functions of a relativistic quantum field theory.  In one-loop computations, it determines the branch on which the logarithms are to be evaluated.  All output expressions of \comm{LoopRefine} observe the $+i\varepsilon$ prescription and are consistent with the analytic conventions of the built-in \emph{Mathematica} functions \comm{Log} and \comm{PolyLog}, which are
\begin{align*}
\text{\comm{Log[x]}}&\longrightarrow \lim_{\varepsilon\rightarrow0^+}\ln(x+i\varepsilon)\,,\enspace\text{and}\\
\text{\comm{PolyLog[2,x]}}&\longrightarrow \lim_{\varepsilon\rightarrow0^+}\text{Li}_2(x-i\varepsilon)\,.
\end{align*}
Because {\packageX} assumes real external invariants and internal masses, almost all analytic formulae can be expressed compactly in terms of the built-in functions.

Whenever \comm{LoopRefine} generates a logarithm containing the ratio of two internal masses, the ratio may be flipped to bring the logarithm to `canonical form', \emph{e.g.}
\begin{equation}
\text{\comm{Log\Big[\frac{m1^2}{m0^2}\Big]} $\longrightarrow$ \comm{-Log\Big[\frac{m0^2}{m1^2}\Big]}\,.}
\end{equation}
Since internal masses are assumed to be positive real, this is allowed, and helps to keep the logarithmic parts compact.  

In the course of reduction, regardless of whether the final expression is divergent or finite, multiple logarithms of ratios of several scales with the 't Hooft parameter $\mu^2$ are typically generated, \emph{e.g.}
\begin{equation}\label{eq:exampleLog}
\text{\comm{a\, Log\Big[\frac{\mu R^2}{-s}\Big] + b\,Log\Big[\frac{\mu R^2}{m^2}\Big] + c\,Log\Big[\frac{\mu R^2}{m^2-s}\Big]}}\,.
\end{equation}  
It is found that by consistently keeping $\mu^2$ in the numerator, the $+i\varepsilon$ \emph{prescription} is always observed -- even when the other scales are external invariants that may become time-like.  The $\mu^2$ from each logarithm are brought into a single logarithm by forming the ratio with a variable that is known to be positive (which were recorded at {\sc step 1}):
\begin{multline}
(\ref{eq:exampleLog}) = \text{\comm{a\, Log\Big[\frac{m^2}{-s}\Big] + (a+b+c)\,Log\Big[\frac{\mu R^2}{m^2}\Big]}}\\
\text{\comm{ +\, c\,Log\Big[\frac{m^2}{m^2-s}\Big]}}\,.
\end{multline}
That way, the coefficient---\comm{(a\!+\!b\!+\!c)} in this example---of the $\mu^2$-dependent logarithm always matches that of the $1/\epsilon$ pole elsewhere in the expression, and are grouped before presenting the results.  If the final expression were in fact finite without a $1/\epsilon$ pole, the coefficient would cancel exactly.

In more complicated cases, expressions cannot be given compactly assuming a universal sign for the infinitesimal imaginary part.  In this case, \emph{Mathematica}'s built-in functions \comm{Log} and \comm{PolyLog} are unsuitable.  For this purpose, two new analytic functions are defined in {\packageX} (within \comm{OneLoop`}):
\begin{align*}
\text{\comm{Ln[x,}$a$\comm{]}}&\longrightarrow \lim_{\varepsilon\rightarrow0^+}\ln(x+ia\varepsilon)\,,\enspace\text{and}\\
\text{\comm{DiLog[x,}$a$\comm{]}}&\longrightarrow \lim_{\varepsilon\rightarrow0^+}\text{Li}_2(x+ia\varepsilon)\,.
\end{align*}
The (real part of the) second argument $a$ controls the side of the branch on which these functions evaluate.  A simple example that uses \comm{DiLog} in its output can be found by running 
\begin{equation*}
\text{\comm{LoopRefine[pvC0[0,m^2,s,0,m,m]]}}\,.
\end{equation*}

\section{{\tt Spur}: Computation of traces of Dirac matrices} \label{sec:Spur}
To assist in the evaluation of one-loop integrals with internal (closed) fermion lines, {\packageX} includes the rudimentary function \comm{Spur} (inside the module \comm{Spur`}) to evaluate traces over products of Dirac gamma matrices appearing in numerators.  The function \comm{Projector} helps to handle loop integrals with open fermion lines and is described in the next section.  Because the primary function of {\packageX} is to compute loop integrals, with the computation of traces being a secondary feature, only a cursory description of the algorithms are given in the following two sections.

As with the rest of the algorithms in {\packageX}, the calculation of traces is rule-based at its core, and bears some resemblance to that of a much earlier \emph{Mathematica} package {\sc Tracer}\cite{Jamin:1991dp}.  However there are a number of differences listed below that lead to greater computation speed.

\begin{itemize}
\item  Throughout the evaluation of the trace, expressions can grow very large containing many terms.  Groups of terms are temporarily enclosed within a \comm{List} to prevent the \emph{Mathematica} kernel from automatically simplifying the large expression at each step of the computation process.

\item While more complicated trace formulae such as those for products of numerous gamma matrices are recursive, non-iterative rules are used for simpler tasks such as for collecting $\gamma_5$ and  $\hat{P}_L$/$\hat{P}_R$ within each term.

\item Products of gamma matrices with repeated Lorentz indices (such as $\gamma^\mu \gamma^\nu \gamma^\rho \gamma_\mu$) are related to products with fewer gamma matrices.
With more gamma matrices interposed between contracted matrices, the number of terms in the identity grows.  On account of the cyclic property of the trace, these contraction identities may be applied in one of two directions.  Additional rules are included so as to apply the identity in the direction with fewer number of interposed gamma matrices.

\item Traces that multiply $\gamma_5$ are tagged differently to set it apart from those without it.  This way, rules for computing traces with $\gamma_5$ and those without $\gamma_5$ are separated, and saves some time when the kernel searches for the appropriate rules.
\end{itemize}

When compared to the other \emph{Mathematica} packages {\sc FeynCalc} and {\sc Tracer}, {\packageX} generally gives results around 10 times faster.  As an example, the trace 
\begin{multline}
\Tr\big[(\slashed{k}-\slashed{p}_1-\slashed{p}_2 + m)\gamma^\nu(g_L \hat{P}_L+ g_R \hat{P}_R)(\slashed{k}-\slashed{p}_2 + m)
\\
\gamma^\rho(g_L \hat{P}_L+ g_R \hat{P}_R)(\slashed{k}+m)\gamma^\mu(g_L \hat{P}_L+ g_R \hat{P}_R)
\big]
\end{multline}
was calculated with each package and computation times were recorded (with \comm{Timing}).  The results on a 2.93 GHz Intel i7 processor are:\\[2mm]
\begin{tabular}{p{3cm}l l}
\packageX & 0.096 & s\\
\sc{FeynCalc} 8.2.0 & 1.03 & s\\
\sc{Tracer} 1.1 & 0.81 & s
\end{tabular}\\

Part of the motivation for refining the trace-taking algorithms is due to the inclusion of fermion projectors described in the next section.  When a \comm{Projector} is included inside \comm{Spur}, the number of terms within the trace is increased to an extent that a noticeable slowdown is observed.  However, with the refinements described above, projections onto form factors are nearly instantaneous on a modern computer.

\section{{\tt Projector}: Projection onto fermion form factors} \label{sec:Projector}
{\packageX} does not directly handle expressions involving open fermion chains that are relevant for fermion self energy and form factor calculations.  In order to provide some support for such computations, {\packageX} comes equipped with a set of projectors.  The projectors permit the projection of a loop integral with an open fermion line onto specific form factors functions.

For example, the one-loop expression for the off-shell fermion self-energy function takes the form
\begin{equation}\label{eq:ExProj}
I(\slashed{p}) = \mu^{2\epsilon}\int\frac{d^d k}{(2\pi)^d} \frac{\mathbb{M}(k, p)}{[k^2-m^2][(k+p)^2]}\,,
\end{equation}
where $\mathbb{M}(k, p)$ is a Dirac matrix structure that depends on the integration variable $k$ and external momentum $p$.  Parity conservation and Lorentz covariance allow $I$ to be written in the form
\begin{equation}
I(\slashed{p}) = A(p^2) \slashed{p} + B(p^2) m\,,
\end{equation}
where the form factors $A$ and $B$ depend on Lorentz invariants $p^2$ and $m^2$ only.  By multiplying the appropriate projectors
\begin{equation*}
\mathcal{F}^{[A]}(p,m) = \frac{1}{4 p^2}\slashed{p}\enspace\,\text{and}\enspace \mathcal{F}^{[B]}(p,m) = \frac{1}{4 m^2}
\end{equation*}
with the numerator of (\ref{eq:ExProj}), and taking the trace, the form factors are obtained:
\begin{align*}
A(p^2) &= \mu^{2\epsilon}\int\frac{d^d k}{(2\pi)^d} \frac{\Tr[\mathbb{M}(k, p)\, \mathcal{F}^{[A]}(p,m)]}{[k^2-m^2][(k+p)^2]}\\
B(p^2) &= \mu^{2\epsilon}\int\frac{d^d k}{(2\pi)^d} \frac{\Tr[\mathbb{M}(k, p)\, \mathcal{F}^{[B]}(p,m)]}{[k^2-m^2][(k+p)^2]}\,.
\end{align*}
The trace over the projectors convert the expressions into ordinary tensors integrals that are readily computed with {\packageX}.

A large set of pre-programmed projectors for off-shell self energy functions and on-shell scalar- and vector-vertex functions in various bases (L/R-chiral or Vector/Axial-vector) are available (as \comm{Projector}) to streamline the computation of such integrals.  These projectors are generalizations of those used in \cite{Kinoshita:1992b} for the calculation of lepton anomalous magnetic moments, and in \cite{Czarnecki:1996rx} for dipole moments.  A comprehensive list of available projectors is given in the built-in documentation files.

\section{Crosschecks and further development}
The verification of loop integrals obtained by {\packageX} is divided into two parts: checking the reduction algorithms in Section \ref{sec:LoopRefine}, and checking the basis functions in Table \ref{tab:ListOfCases}.  The reduction routines for $A$ and $B$ coefficient functions and the basis $B_{1\ldots1}$ functions were compared against another (unpublished) computer program developed by Huaike Guo.  The reduction of $C$ functions for \emph{Cases 1, 3, 5} and \emph{6} were checked against explicit formulae for the low rank functions listed in [DD].  Each $C_0$ scalar function was derived by hand and compared against explicit formulae in the literature where they exist \cite{Angel:2013hla, Ellis:2007qk, JorgeRomao:2004a, CabralRosetti:2002qv}.  In cases where they did not exist, the analytic expressions were checked by comparing with the results of numerically integrating the corresponding Feynman parameter representations given in Appendix \ref{sec:FeynPar}.

Finally, as a combined check of the various algorithms in {\packageX}, the following well known physical quantities were computed and verified: $H\rightarrow gg$ and $\gamma\gamma$ standard model decay rates \cite{Djouadi:2005gi}, electron $g-2$, and the neutrino electric and magnetic moments \cite{Giunti:2014ixa}.  Each was found to be in agreement with literature.

There are a number of important limitations of {\packageX}, listed below, that guides its current line of development.
\begin{enumerate}
\item An analytic series expansion of the loop integral in kinematic variables is not generally possible.  Currently, the only available method is to use \emph{Mathematica}'s \comm{Series} on the output of \comm{LoopRefine}.  However, if the result of loop integral contains specially defined function like \comm{pvC0}, then \comm{Series} will not work.  Given that much information about a loop-integral can be gleaned from its expansion, the omission of this feature is most conspicuous.  
\item {\packageX} currently supports loop integrals with up to only three denominator factors.  But, as the number of denominator factors increases, so does the complexity of their analytic forms.  Thus, it would not be so practical to work with such expressions for higher-point functions even if {\packageX} were to provide them.  However, at special kinematic points such as at zero external momenta or at thresholds compact expressions could be obtained.

\item Gamma-5 is implemented naively in dimensional regularization.  This means that the VVA or AAA three-point functions may not automatically satisfy Ward identities appropriate to the physical problem.  However, the versatility of {\packageX} makes it easy to apply Adler's method \cite{Adler:method} (see also \cite{Jegerlehner:2000dz}) to enforce the Ward identities.

\item Loop integrals with open fermion chains are not directly supported.  As explained in Section \ref{sec:Projector}, there is no way to input an open string of Dirac matrices.  Instead, the computation of fermion form factors can be done by projecting out the needed form factors.
\end{enumerate}

\begin{acknowledgements}
I express my gratitude to the members of the \emph{Mathematica} StackExchange community for providing countless answers to my questions regarding the technical aspects of \emph{Mathematica} and \emph{Wolfram Workbench}.  I thank Ansgar Denner for clarifying discussions regarding the reduction methods and for providing helpful information regarding the literature.  Numerous colleagues have provided useful feedback: Huaike Guo for cross-checking several results during the early stages of the development, and Xunjie Xu for encouraging me to refine the trace-taking algorithms.  

I also acknowledge my debt to beta tester Johannes Welter for cross-checking the fermion form factor projectors, and to beta testers Juri Smirnov and Michael Duerr for identifying bugs.  Special thanks goes to Michael Duerr for meticulously hand-checking various results of the reduction algorithms, and for suggesting improvements to the user interface, the tutorial, and the accompanying documentation pages.
\end{acknowledgements}

\appendix
\begin{table}[b]
\centering
\begin{tabular}{lll}
\parbox[c]{2.5cm}{ Quantity} & \parbox[c]{4.0cm}{Convention} \\[3mm]
\hline
Metric signature & $g_{\mu\nu} = \text{diag}(+,-,-,-)$ \\[1mm]
Spacetime dimension & $d = 4-2\epsilon$\\[1mm]
Dirac matrix commutator & $\sigma_{\mu\nu}= \frac{i}{2}[\gamma_\mu,\gamma_\nu]$ \\[1mm]
Fifth gamma matrix & $\gamma_5 = i \gamma^0 \gamma^1 \gamma^2 \gamma^3$\\[1mm]
Chiral projectors & $\hat{P}_L = \frac{1}{2}(1-\gamma_5)$,\,$\hat{P}_R = \frac{1}{2}(1+\gamma_5)$\\[1mm]
Levi-civita symbol & $\epsilon^{0123} = +1$
\end{tabular}
\caption{Conventions for spacetime quantities}\label{tab:Conventions}
\end{table}
\begin{table}
\centering
\begin{tabular}{rc}
\parbox[c]{3.5cm}{Function} & \parbox[c]{3.0cm}{Diagram} \\[2mm]
\hline\\
$A_0(m_0)$ & \parbox{1.2cm}{\includegraphics[width=1.2cm]{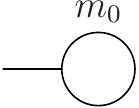}} \\[7mm]
\begin{tabular}{r}
$B_0(p^2, m_0, m_1),$ \\ and\enspace $b_0^\xi(p^2, m_0, 0)$
\end{tabular} & \parbox{2.04cm}{\includegraphics[width=2.04cm]{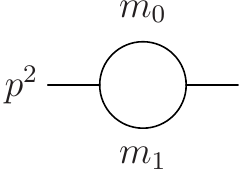}}\\[7mm]
\begin{tabular}{r}
$C_0(p_1^2,p_2^2,q^2, m_2, m_1, m_0)\,,$ \\[2mm] $q^2 = (p_2-p_1)^2$
\end{tabular}
 & \parbox{2.28cm}{\includegraphics[width=2.28cm]{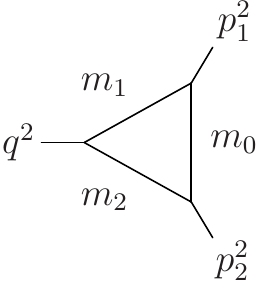}} 
\end{tabular}
\caption{Conventions for the arguments of the Passarino-Veltman functions}\label{tab:PVConventions}
\end{table}
\section{Conventions}\label{app:conventions}
For reference, the conventions for spacetime quantities are summarized in Table \ref{tab:Conventions}.  Conventions for the Passarino-Veltman functions are displayed in Table \ref{tab:PVConventions}.  Note that a slightly-unconventional ordering and form for the arguments of the Passarino-Veltman functions is taken.  However, this choice makes the invariance property under their pairwise interchange clear:
\begin{align}\label{eq:invarianceOfC0}
\nonumber C_0(p_1^2,p_2^2,q^2, m_2, m_1, m_0)\\
\nonumber &\hspace{-2cm}= C_0(p_2^2,p_1^2,q^2, m_1, m_2, m_0)\\
 &\hspace{-2cm}= C_0(q^2,p_2^2,p_1^2, m_0, m_1, m_2)\,.
\end{align}

\section{Feynman parameter integral representations of Passarino-Veltman coefficient functions}\label{sec:FeynPar}
In this section, the Feynman parameter integral representation of the Passarino-Veltman coefficient one, two, and three point functions are given.  They are obtained \cite{Davydychev:1991va} by writing tensor integrals as derivatives of the integral representation of the corresponding scalar integral with respect to external momenta, and then matching the result to the respective covariant tensor decomposition.
\begin{equation}\label{eq:FeynParA}
A_{\underbrace{0\ldots0}_{2r}}(m_0)=(4\pi\mu^2)^\epsilon \frac{(-1)^{1+r}}{2^r}\Gamma(-1+\epsilon-r) m_0^{1-\epsilon+r}
\end{equation}
\begin{multline}
\label{eq:FeynParB}
B_{\underbrace{0\ldots0}_{2r}\underbrace{1\ldots1}_{n}}(p^2;m_0,m_1) = (4\pi\mu^2)^\epsilon \frac{(-1)^{2+r+n}}{2^r}\Gamma(\epsilon-r)\\
\times\int_0^1 \frac{dx\, x^n}{\big(p^2 x^2+(-p^2+m_1^2-m_0^2)x + m_0^2 - i\varepsilon\big)^{\epsilon-r}}
\end{multline}
\begin{multline}\label{eq:FeynParBaux}
b^\xi_{\underbrace{0\ldots0}_{2r}\underbrace{1\ldots1}_{n}}(p^2;m)=(4\pi\mu^2)^\epsilon \frac{(-1)^{3+r+n}}{2^r}\Gamma(1+\epsilon-r)\\
\times\int_0^1 \frac{dx \,x^{n} (1-x)}{\big(p^2 x^2+(-p^2+m^2)x - i\varepsilon\big)^{1+\epsilon-r}}
\end{multline}
\begin{widetext}
\begin{multline}\label{eq:FeynParC}
C_{\underbrace{0\ldots0}_{2r}\underbrace{1\ldots1}_{n_1}\underbrace{2\ldots2}_{n_2}}(p_1^2,p_2^2,q^2;m_2,m_1,m_0) = (4\pi\mu^2)^\epsilon \frac{(-1)^{3+r+n_1+n_2}}{2^r}\Gamma(1+\epsilon-r)\\
\times\int_0^1 dy\int_0^{1-y} dz \, y^{n_1} z^{n_2} \big[p_1^2 y^2 + p_2^2 z^2 + (-q^2+p_1^2+p_2^2)yz + (-p_1^2+m_1^2-m_0^2)y+(-p_2^2+m_2^2-m_0^2)z+m_0^2-i\varepsilon\big]^{-1-\epsilon+r}
\end{multline}
The coefficient $C$ function exhibits an invariance under the simultaneous interchange of indices $n_1\leftrightarrow n_2$, external momenta $p_1^2 \leftrightarrow p_2^2$ and internal masses $m_1\leftrightarrow m_2$,
\begin{equation}\label{eq:invarianceOfC}
C_{\underbrace{0\ldots0}_{2r}\underbrace{1\ldots1}_{n_1}\underbrace{2\ldots2}_{n_2}}(p_1^2,p_2^2,q^2;m_2,m_1,m_0)=C_{\underbrace{0\ldots0}_{2r}\underbrace{1\ldots1}_{n_2}\underbrace{2\ldots2}_{n_1}}(p_2^2,p_1^2,q^2;m_1,m_2,m_0)
\end{equation}
and is frequently employed during its reduction in the most general kinematic case ($\det Z\neq0$).

In certain reduction formulae of $C$-functions, some terms are multiplied by $\epsilon$, which in the $\epsilon\rightarrow0$ limit, pick up the UV-divergent parts of the coefficient functions in those terms. The UV divergent parts are readily obtained from the integral representation.  They are controlled by the leading gamma function which for large enough $r$ develops a $1/\epsilon$ pole as $\epsilon\rightarrow0$.  When $r$ is large, the integrand becomes polynomial in the Feynman parameters and are readily integrated with the help of the multinomial theorem.   The needed UV-divergent parts are those of the $B$ and $C$ functions, shown below.
\begin{equation}
B_{\underbrace{0\ldots0}_{2r}\underbrace{1\ldots1}_{n}}(p^2;m_0,m_1)\Big|_{\substack{\text{UV-}\\\text{Div.}}} = \frac{(-1)^{n}}{2^r r!} \sum_{k_1+k_2+k_3=r} \binom{r}{k_1,\,k_2,\,k_3}\frac{a^{k_1} b^{k_2} c^{k_3}}{2k_1+k_2+n+1}\frac{1}{\bar\epsilon}\,,
\end{equation}
where $a=p^2$, $b=-p^2+m_1^2-m_0^2$, and $c=m_0^2$, are polynomial coefficients of the integrand in (\ref{eq:FeynParB}).
\begin{multline}
C_{\underbrace{0\ldots0}_{2r}\underbrace{1\ldots1}_{n_1}\underbrace{2\ldots2}_{n_2}}(p_1^2,p_2^2,q^2;m_2,m_1,m_0)\Big|_{\substack{\text{UV-}\\\text{Div.}}}\\
= \frac{(-1)^{n_1+n_2}}{2^r (r-1)!}\sum_{k_1+\ldots+k_6=r-1}\binom{r-1}{k_1,\ldots,k_6} a^{k_1} b^{k_2} c^{k_3} d^{k_4} e^{k_5} f^{k_6} \frac{(2k_1+k_3+k_4+n_1)!(2k_2+k_3+k_5+n_2)!}{(2k_1+2k_2 +2k_3 +k_4 +k_5+n_1+n_2+2)!}\frac{1}{\bar\epsilon}\,,
\end{multline}
where $a$, $b$, $c$, $d$, $e$, and $f$ are polynomial coefficients of the integrand in (\ref{eq:FeynParC}) in the order displayed.

\section{Derivation of reduction formulae for $C$ functions \emph{Case 2}}\label{app:SpecialCaseDerivation}
The derivation of the first equation in (\ref{eq:ez6A}) begins with the Feynman parameter representation of the coefficient $C$ function,
\begin{multline}\label{eq:startDerivation1}
C_{\underbrace{0\ldots0}_{2r}\underbrace{1\ldots1}_{n_1}\underbrace{2\ldots2}_{n_2}}(m_0^2,s,m_2^2;m_2,0,m_0) = (4\pi\mu^2)^\epsilon \frac{(-1)^{3+r+n_1+n_2}}{2^r}\Gamma(1+\epsilon-r)\\
\times\int_0^1 dy\int_0^{1-y} dz \, y^{n_1} z^{n_2} \big[m_0^2 y^2 + s z^2 + (-m_2^2 +m_0^2 +s)y z + (-m_0^2 + m_2^2 - s)z+m_0^2 -i\varepsilon\big]^{-1-\epsilon+r}\,.
\end{multline}
Upon making a change of integration variables $y = 1 - y'$ and  $z = y'z'$, the nested integrals are factored:
\begin{equation}
\text{integrals}=\int_0^1 dy'\, y'^{-1+n_2-2\epsilon+2r}(1-y')^{n_1}\,\int_0^1 dz'\, z'^{n_2}\big[s z'^2 + (-s+m_2^2-m_0^2)z' + m_0^2 - i\varepsilon\big]^{-1-\epsilon+r}\,.
\end{equation}
The $y'$ integral gives the Euler Beta function, while the $z'$ integral is identified as the integral representation of coefficient $B$-function (\ref{eq:FeynParB}).
\begin{equation}
C_{\underbrace{0\ldots0}_{2r}\underbrace{1\ldots1}_{n_1}\underbrace{2\ldots2}_{n_2}}(m_0^2,s,m_2^2;m_2,0,m_0) = \frac{(-1)^{n_1}}{2} \text{B}(n_2-2\epsilon+2r,n_1+1) B_{\underbrace{0\ldots0}_{\mathclap{2r-2}}\underbrace{1\ldots1}_{n_2}}(s;m_0,m_2)
\end{equation}
As long as one of $n_2$ or $r$ is non-zero, the Beta function is finite, and its expansion to $\mathcal{O}(\epsilon)$ may be inserted yielding the first equation in (\ref{eq:ez6A}).  

On the other hand, if $n_2=r=0$, the Beta function develops a $1/\epsilon$ pole, so that to $\mathcal{O}(\epsilon)$, \\ \mbox{$B(-2\epsilon,\,n_1+1)=\frac{-1}{2\epsilon}-H_{n_1}-\epsilon\big(H_{n_1}^2-H_{n_1}^{(2)}\big)$}.  In this case, (\ref{eq:startDerivation1}) is written as
\begin{multline}
C_{\underbrace{1\ldots1}_{n_1}}(m_0^2,s,m_2^2;m_2,0,m_0) = (4\pi\mu^2)^\epsilon (-1)^{n_1} \Gamma(1+\epsilon) \frac{1}{2\epsilon}\int_0^1 dz' \big(s z'^2 + z'(-s+m_2^2-m_0^2)+m_0^2 - i\varepsilon\big)^{-1-\epsilon}\\
+(4\pi\mu^2)^\epsilon (-1)^{n_1}\Gamma(1+\epsilon)\big(H_{n_1}+\epsilon(H_{n_1}^2-H_{n_1}^{(2)})\big)\int_0^1 dz' \big(s z'^2 + z'(-s+m_2^2-m_0^2)+m_0^2 - i\varepsilon\big)^{-1-\epsilon}\,.
\end{multline}
While the $z'$ integral in the second line can be identified with the integral representation of the coefficient $B$ function, the first line is identified\footnote{see http://qcdloop.fnal.gov/tridiv6.pdf} as the integral representation of the scalar function $C_0(m_0^2,s,m_2^2;m_2,0,m_0)$ classified by Ellis and Zanderighi \cite{Ellis:2007qk} as IR-divergent triangle 6.  These identifications lead to the second equation of (\ref{eq:ez6A}).

If the off-shell momentum $s$ is in the third argument, the derivation starts with the change of variables $z=1-y-x$ in (\ref{eq:FeynParC}) followed by an interchange of the $x$ and $y$ integrals to give
\begin{multline}
C_{\underbrace{0\ldots0}_{2r}\underbrace{1\ldots1}_{n_1}\underbrace{2\ldots2}_{n_2}}(m_2^2,m_0^2,s;m_0,m_2,0) = (4\pi\mu^2)^\epsilon \frac{(-1)^{3+r+n_1+n_2}}{2^r}\Gamma(1+\epsilon-r)\\
\times\int_0^1 dx\int_0^{1-x} dy \, y^{n_1} (1-x-y)^{n_2} \big[m_0^2 x^2 + s y^2 + (s+m_0^2-m_2^2)x y - 2m_0^2 x + (-s-m_0^2 +m_2^2)y +m_0^2 -i\varepsilon\big]^{-1-\epsilon+r}.
\end{multline}
The nested integrals are factored by making a further change of variables $x=1-x'$ and $y = y' x'$ to give
\begin{align}
\text{integrals} &= \int_0^1 dx'\, x'^{n_1+n_2+2r-1-2\epsilon}\,\int_0^1 dy'\,y'^{n_1}(1-y')^{n_2}\big[s y'^2 + (-s+m_2^2 -m_0^2)y' +m_0^2 -i\varepsilon\big]^{-1-\epsilon+r}\,.
\intertext{The $x'$ integral is straightforward.  The $y'$ integral can be brought to a recognizable form after expanding the factor $(1-y')^{n_2}$ as a binomial series}
&=\frac{1}{n_1+n_2+2r-2\epsilon}\sum_{k=0}^{n_2}\binom{n_2}{k}(-1)^k \int_0^1 dy'\, y'^{n_1+k} \big[s y'^2+(-s+m_2^2-m_0^2)y' + m_0^2-i\varepsilon\big]^{-1-\epsilon+r}\,.
\end{align}
The $y'$ integral is now identified as the integral representation of the coefficient $B$ function, yielding (\ref{eq:ez6B}).

\section{Derivation of reduction formulae for $C$ functions \emph{Case 4}}\label{sec:DerivationOfCase4}
Two cases are distinguished for \emph{Case 4} ($\det Z = 0$, $\tilde X_{0j} = 0$) depending on whether $p_2^2$ is vanishing.  Although the steps below leading to (\ref{eq:reductionAtThres1}) and (\ref{eq:reductionAtThres2}) appear complicated, they essentially follow that of \cite{Hooft:1978xw} for the evaluation of the scalar function $C_0$.  Beginning with the integral representation (\ref{eq:FeynParC}), a change of integration variables $y=1-y'$ brings the Feynman integrals to the form
\begin{equation}\label{eq:afterInitialChange}
\text{integrals} = \int_0^1 dy'\int_0^{y'} dz \, (1-y')^{n_1} z^{n_2} \big[a\, y'^2 + b\,z^2 + c\, y'z  + d\, y' + e\,z+f\big]^{-1-\epsilon+r}
\end{equation}
where $a=p_1^2$, $b=p_2^2$, $c=q^2-p_1^2-p_2^2$, $d=-p_1^2+m_0^2-m_1^2$, $e=p_1^2-q^2-m_0^2+m_2^2$, and $f=m_1^2-i\varepsilon$.

Under the assumption that $p_2^2 \neq 0$, a second change of variables is made $z = z' + \alpha y'$, with $\alpha=\frac{-c}{2b}$ chosen to make the coefficient of $y'^2$ in square brackets vanish.
\begin{equation}
\text{integrals} = \int_0^1 dy' \int_{-\alpha y}^{(1-\alpha)y}dz'\,(1-y')^{n_1}(z'+\alpha y')^{n_2}\big[b z'^2 + (c+2b\alpha) y'z' +(d+e\alpha)y'+e\,z'+f\big]^{-1-\epsilon+r}
\end{equation}
The choice for $\alpha$ implies that $c+2b\alpha$ vanishes, and the kinematic relations $\det Z = \tilde X_{0j}=0$ imply that $d+e\alpha$ vanishes, yielding
\begin{equation}
\text{integrals} = \int_0^1 dy \int_{-\alpha y}^{(1-\alpha)y}dz\,(1-y)^{n_1}(z+\alpha y)^{n_2}\big[b z^2 +e\,z+f\big]^{-1-\epsilon+r}\,,
\end{equation}
where the primes have been omitted.  The binomial theorem is applied to the factor $(z+\alpha y)^{n_2} = \sum_j \binom{n_2}{j}\alpha^{n_2-j}y^{n_2-j}z^j$, and the order of integrations is interchanged so that
\begin{equation}\label{eq:Case4IntermediateStep}
\text{integrals} = \sum_{j=0}^{n_2}\binom{n_2}{j}\alpha^{n_2-j}\Big[\int_0^{1-\alpha} dz \int_{z/(1-\alpha)}^1 dy - \int_0^{-\alpha} dz \int_{-z/\alpha}^1 dy \Big](1-y)^{n_1}y^{n_2-j}z^{j}\big[b z^2 +e\,z+f\big]^{-1-\epsilon+r}\,.
\end{equation}
The $y$ integrals in both terms yield terminating hypergeometric series most compactly written in terms of the incomplete Beta function:
\begin{align}\nonumber
\int_X^1 dy (1-y)^{n_1} y^{n_2-j} &= \frac{n_1!(n_2-j)!}{(n_1+n_2-j+1)!}-\text{B}_X(n_2-j+1, n_1+1)\\
\label{eq:incompleteBeta} &= \frac{n_1!(n_2-j)!}{(n_1+n_2-j+1)!} - \sum_{k=0}^{n_1} \frac{(-1)^k X^{n_2-j+k+1}}{(n_2-j+k+1)}\binom{n_1}{k}\,.
\end{align}
Since the first term of (\ref{eq:incompleteBeta}) is common to both integrations in (\ref{eq:Case4IntermediateStep}), they are combined to yield a total of three terms:
\begin{multline}
\text{integrals} = \sum_{j=0}^{n_2}\binom{n_2}{j}\alpha^{n_2-j}\Big[\frac{n_1!(n_2-j)!}{(n_1+n_2-j+1)!} \int_{-\alpha}^{1-\alpha} dz \frac{z^{n_2+j}}{(b z^2 + ez+f)^{1+\epsilon-r}}\\
-\int_0^{1-\alpha} dz \frac{\text{B}_{z/(1-\alpha)}(n_2-j+1,n_1+1)\, z^j}{(b z^2 + ez+f)^{1+\epsilon-r}} + \int_0^{-\alpha} dz \frac{\text{B}_{-z/\alpha}(n_2-j+1,n_1+1)\,z^j}{(b z^2 + ez+f)^{1+\epsilon-r}}\Big]
\end{multline}
A change of integration variables is carried out in each term to stretch their ranges to $0\rightarrow 1$:  In the first integral, $z=z'-\alpha$, in the second integral $z=(1-\alpha)z'$, and in the third integral $z=-\alpha z'$.  Consequently, the polynomials $b z^2 + e z+f$ take the shape of integrands for the $B$ functions\footnote{These quadratic polynomials may be recognized as the `pinch functions' originating from the three cut channels of the triangle graph.}:
\begin{equation*}
\begin{array}{p{2.5cm}cl}
\text{First term:}& p_2^2 z'^2 + (-p_2^2 +m_2^2-m_0^2)z' + m_0^2 -i\varepsilon &:=P_2(z')\\
\text{Second term:}& q^2 z'^2 + (-q^2 +m_2^2-m_1^2)z' + m_1^2 -i\varepsilon &:= P_{12}(z')\\
\text{Third term:}& p_1^2 z'^2 + (-p_1^2 +m_0^2-m_1^2)z' + m_1^2 -i\varepsilon &:= P_1(z')
\end{array}
\end{equation*}
Upon inserting the series representation of the incomplete Beta function (\ref{eq:incompleteBeta}) the result is (after dropping the primes on $z$)
\begin{multline}
\text{integrals} = \sum_{j=0}^{n_2} \binom{n_2}{j}\alpha^{n_2-j}\bigg\{\frac{n_1!(n_2-j)!}{(n_1+n_2-j+1)!} \int_0^1 dz (z-\alpha)^j P_2(z)^{-1-\epsilon+r}\\
+ \sum_{k=0}^{n_1} \frac{(-1)^{k}}{n_2-j+k+1}\binom{n_1}{k}\Big[-(1-\alpha)^{j+1}\int_0^1 dz\, z^{n_2+k+1} P_{12}(z)^{-1-\epsilon+r}+(-\alpha)^{j+1} \int_0^1 dz\, z^{n_2+k+1}P_1(z)^{-1-\epsilon+r}\Big]
\bigg\}
\end{multline}
In the first term, the binomial theorem is applied to $(z-\alpha)^j=\sum_k\binom{j}{k}(-\alpha)^{j-k}z^k$, and the three $z$ integrals are finally identified as integral representations of the coefficient $B$ functions upon which (\ref{eq:reductionAtThres1}) is obtained.

If $p_2^2=0$, equation (\ref{eq:reductionAtThres1}) breaks down and another formulae is needed.  In this case, $\det Z=0$ implies $p_1^2 = q^2$ and $\tilde X_{0j}=0$ implies $m_0=m_2$ provided $p_1^2 \neq 0$.  With these relations, the integrals in (\ref{eq:afterInitialChange}) are already factored:
\begin{equation}
\text{integrals}=\int_0^1 dy' \int_0^{y'} dz \, (1-y')^{n_1} z^{n_2} \big[p_1^2 y'^2 +(-p_1^2+m_0^2-m_1^2)y'+m_1^2-i\varepsilon\big]^{-1-\epsilon+r}\,.
\end{equation}
The $z$ integration gives a factor $1/(n_2+1)$, and the factor $(1-y')^{n_1}=\sum_k \binom{n_1}{k}(-y)^k$ is expressed as a binomial series.
\begin{equation}
\text{integrals}=\frac{1}{n_2+1}\sum_{k=0}^{n_1}\binom{n_1}{k}(-1)^k\int_0^1 dy' y'^{n_2+k+1}\big[p_1^2 y'^2 +(-p_1^2+m_0^2-m_1^2)y'+m_1^2-i\varepsilon\big]^{-1-\epsilon+r}\,.
\end{equation}
Eqn (\ref{eq:reductionAtThres2}) is obtained after identifying the $y'$ integration as the integral representation of the coefficient $B$ function.  If all external invariants are vanishing $p_1^2 = p_2^2 = q^2 = 0$ then neither (\ref{eq:reductionAtThres1}) nor (\ref{eq:reductionAtThres2}) are valid, and \emph{Case 5} is needed.
\end{widetext}

\section{Coefficient $B$ functions with $r=-1$}\label{App:algB}
The two new reduction algorithms for $C$ functions (\emph{Cases 2} and \emph{4}) require extending the set of basis functions to include $B$ functions in which the index $r$ in (\ref{eq:FeynParB}) is continued to $-1$.  In a certain sense, these new reduction formulae may be closely related to those in \cite{Duplancic:2003tv}.  There, the authors present different reduction formulae for coefficient functions which likewise require extending the set of basis functions to scalar functions with repeated propagators.

A set of explicit expressions for the general case and at singular points is constructed and included in the {\packageX} source file (see Table \ref{tab:ListOfCases}).  The integration is straightforward in most cases.  The functions are UV-finite for all $n\geq0$, but with many kinematic configurations developing IR-divergent $1/\epsilon$ poles.  

However, there are three kinematic cases, all corresponding to physical threshold for which the Feynman parameter integral nominally diverges even for finite but infinitesimal $\epsilon$.  To handle these cases, $\epsilon$ is taken sufficiently large and negative so that the integral converges, and then analytically continued to $\epsilon\rightarrow0$.  The results of these integrations are given below:
\begin{align}\nonumber\label{eq:algB1}
B_{\underbrace{0\ldots0}_{r=-1} \underbrace{1\ldots 1}_{0}}(m_1^2;0,m_1) \\
\nonumber&\hspace{-1cm}= \frac{-2}{m_1^2} \big(\frac{4\pi \mu^2}{m_1^2}\big)^\epsilon\, \Gamma(1+\epsilon)\int_0^1 dx\, x^{-2-2\epsilon}\\
&\hspace{-1cm}= \frac{2}{m_1^2}
\end{align}
\begin{align}\nonumber\label{eq:algB2}
B_{\underbrace{0\ldots0}_{r=-1} \underbrace{1\ldots 1}_{n}}(m_0^2;m_0,0)\\
\nonumber&\hspace{-3cm} = \frac{2}{m_0^2} \big(\frac{4\pi \mu^2}{m_0^2}\big)^\epsilon (-1)^{n+1}\Gamma(1+\epsilon)\int_0^1 dx\, x^n (x-1)^{-2-2\epsilon}\\
&\hspace{-3cm}= \begin{cases}\frac{(-1)^{n+1}}{m_0^2} n \Big(\frac{1}{\bar\epsilon}+\ln\big(\frac{\mu^2}{m_0^2}\big)+2 H_{n-1}-2\Big), & n \geq 1 \\ \frac{2}{m_0^2}, & n=0\end{cases}
\end{align}
\begin{multline*}
B_{\underbrace{0\ldots0}_{r=-1} \underbrace{1\ldots 1}_{n}}\big((m_0\!+\!m_1)^2;m_0,m_1\big) = \frac{2(-1)^{n+1}}{(m_0\!+\!m_1)^2}\\
 \times \Big(\frac{4\pi \mu^2}{(m_0\!+\!m_1)^2}\Big)^\epsilon \Gamma(1+\epsilon) \int_0^1 dx \frac{x^n}{\big[\big(x-x_+\big)^2\big]^{1+\epsilon}}
\end{multline*}
\begin{multline}\label{eq:algB3}
= \frac{2(-1)^{n+1}}{(m_0\!+\!m_1^2)}\Big[\sum_{k=0}^{n-2}\frac{x_+^{n-2-k}}{k+1}+n\,x_+^{n-1}\ln\big({\textstyle 1 - \frac{1}{x_+}}\big)\\
 -\frac{1}{1-x_+}-\frac{\delta_{n,0}}{x_+}\Big],\qquad x_+ = \frac{m_0}{m_0-m_1}
\end{multline}
Among these integrals, only (\ref{eq:algB3}) gives numerical results that are related to limiting values as threshold is reached: the real part of (\ref{eq:algB3}) corresponds to the limiting value of $\text{Re}\, B(s;m_0,m_1)$ when approached from above threshold, and the imaginary part corresponds to the limiting value of $\text{Im}\, B(s;m_0,m_1)$ when approached below threshold.

That the integrals (\ref{eq:algB1}-\ref{eq:algB3}) give numerical results that do not match their limiting values as threshold is reached are not of any concern.  The results above should be viewed as ill-defined divergent integrals arising at intermediate stages in the reduction of coefficient $C$ functions.  They serve to facilitate the analytic cancellation of these integrals at the end of a physically meaningful computation, such as for the electromagnetic contribution to the electron anomalous magnetic moment.

\bibliography{packageX.bib}

\end{document}